\documentclass[aps,prl,reprint,groupedaddress,showpacs, showkeys]{revtex4-1}

\usepackage{amssymb,amsfonts,amsmath}
\usepackage{amsmath}
\usepackage{amsfonts}
\usepackage{amssymb}

\usepackage[dvips]{graphicx}
\usepackage{pdfpages}

\newcommand{\Rbar}{\bar{R}}
\newcommand{\xbar}{\bar{x}}
\newcommand{\sign}{\textrm{ sgn}}
\newcommand{\half}{\frac{1}{2}}

\newcommand{\boldA}{\mathbf{A}}

\newcommand{\suppsecfps}{S2}  
\newcommand{\suppsecstab}{S3}  
\newcommand{\suppsecdata}{S4}  
\newcommand{\suppsecpert}{S5}  
\newcommand{\suppsecnumerics}{S6}  
\newcommand{\suppsecdelay}{S7}  

\newcommand{\suppfigflow}{S2}  
\newcommand{\suppfignumerics}{S5}  
\newcommand{\suppfigxvst}{S6}  
\newcommand{\suppfigdelay}{S7}  

\begin{document}


\title{A mathematical model of social group competition with application to the growth of religious non-affiliation}


\author{Daniel M. Abrams}
\author{Haley A. Yaple}
\affiliation{Department of Engineering Sciences and Applied Mathematics, Northwestern University, Evanston, Illinois 60208, USA}
\author{Richard J. Wiener}
\affiliation{Research Corporation for Science Advancement, Tucson, Arizona 85712, USA}
\affiliation{Department of Physics, University of Arizona, Tucson, Arizona 85721, USA}


\date{\today}

\begin{abstract}
When groups compete for members, the resulting dynamics of human social activity may be understandable with simple mathematical models.  Here, we apply techniques from dynamical systems and perturbation theory to analyze a theoretical framework for the growth and decline of competing social groups.  We present a new treatment of the competition for adherents between religious and irreligious segments of modern secular societies and compile a new international data set tracking the growth of religious non-affiliation. Data suggest a particular case of our general growth law, leading to clear predictions about possible future trends in society.
\end{abstract}


\maketitle

The tools of statistical mechanics and nonlinear dynamics have been used successfully in the past to analyze models of social phenomena ranging from language choice \cite{abrams03} to political party affiliation \cite{bennaim05} to war \cite{ispolatov96} and peace \cite{zhao09}.  In this work, we focus on social systems comprised of two mutually exclusive groups in competition for members \cite{weidlich00, epstein97, krapivsky03, benczik08, holme06, durrett05}.  We compile and analyze a new data set quantifying the declining rates of religious affiliation in a variety of regions worldwide and present a theory to explain this trend.

People claiming no religious affiliation constitute the fastest growing religious minority in many countries throughout the world\cite{zuckerman07}.  Americans without religious affiliation comprise the only religious group growing in all 50 states; in 2008 those claiming no religion rose to 15 percent nationwide, with a maximum in Vermont at 34 percent\cite{kosmin09}.  In the Netherlands nearly half the population is religiously unaffiliated\cite{NLstats}.  Here we use a minimal model of competition for members between social groups to explain historical census data on the growth of religious non-affiliation in 85 regions around the world.  According to the model, a single parameter quantifying the perceived utility of adhering to a religion determines whether the unaffiliated group will grow in a society.  The model predicts that for societies in which the perceived utility of not adhering is greater than the utility of adhering, religion will be driven toward extinction.

\section{Model}

We begin by idealizing a society as partitioned into two mutually exclusive social groups, $X$ and $Y$, the unaffiliated and those who adhere to a religion.   We assume the attractiveness of a group increases with the number of members, which is consistent with research on social conformity\cite{tanford84, latane96, latane81, hogg90}.  We further assume that attractiveness also increases with the perceived utility of the group, a quantity encompassing many factors including the social, economic, political and security benefits derived from membership as well as spiritual or moral consonance with a group.  Then a simple model of the dynamics of conversion is given by\cite{abrams03}
\begin{equation} \label{eq1}
	\frac{dx}{dt} = y P_{yx}(x, u_x) - x P_{xy}(x, u_x)
\end{equation}
where $P_{yx}(x, u_x)$ is the probability, per unit of time, that an individual converts from $Y$ to $X$, $x$ is the fraction of the population adhering to $X$ at time $t$, $0 \le u_x \le 1$ is a measure of $X$'s perceived utility, and $y$ and $u_y$ are complementary fractions to $x$ and $u_x$.  We require $P_{xy}(x, u_x) = P_{yx}(1 - x, 1 - u_x)$ to obtain symmetry under exchange of $x$ and $y$ and $P_{yx}(x, 0) = P_{yx}(0, u_x) = 0$ to capture the idea that no one will switch to a group with no utility or adherents.  The assumptions regarding the attractiveness of a social group also imply that $P_{yx}$ is smooth and monotonically increasing in both arguments. Under these assumptions, for generic $P_{yx}(x, u_x)$ Eq.~\eqref{eq1} has at most three fixed points, the stability of which depends on the details of $P_{yx}$ (see Supplementary Material Section \suppsecfps).

A functional form for the transition probabilities consistent with the minimal assumptions of the model is $P_{yx}(x,u_x) = c x^a u_x$, where $c$ and $a$ are constants that scale time and determine the relative importance of $x$ and $u_x$ in attracting converts, respectively.  (Supplementary Figure \suppfigflow~illustrates the structure of the fixed points for this case.)  If $a > 1$ there are three fixed points, one each at $x = 0$ and $x = 1$, which are stable, and one at $0 < x < 1$, which is unstable.  For $a < 1$ the stability of these fixed points is reversed.  For the boundary at $a = 1$, there are only two fixed points, one of which is stable and the other unstable (see Supplementary Material Section \suppsecstab).

\begin{figure}[htp]
\includegraphics[width=8 cm,clip]{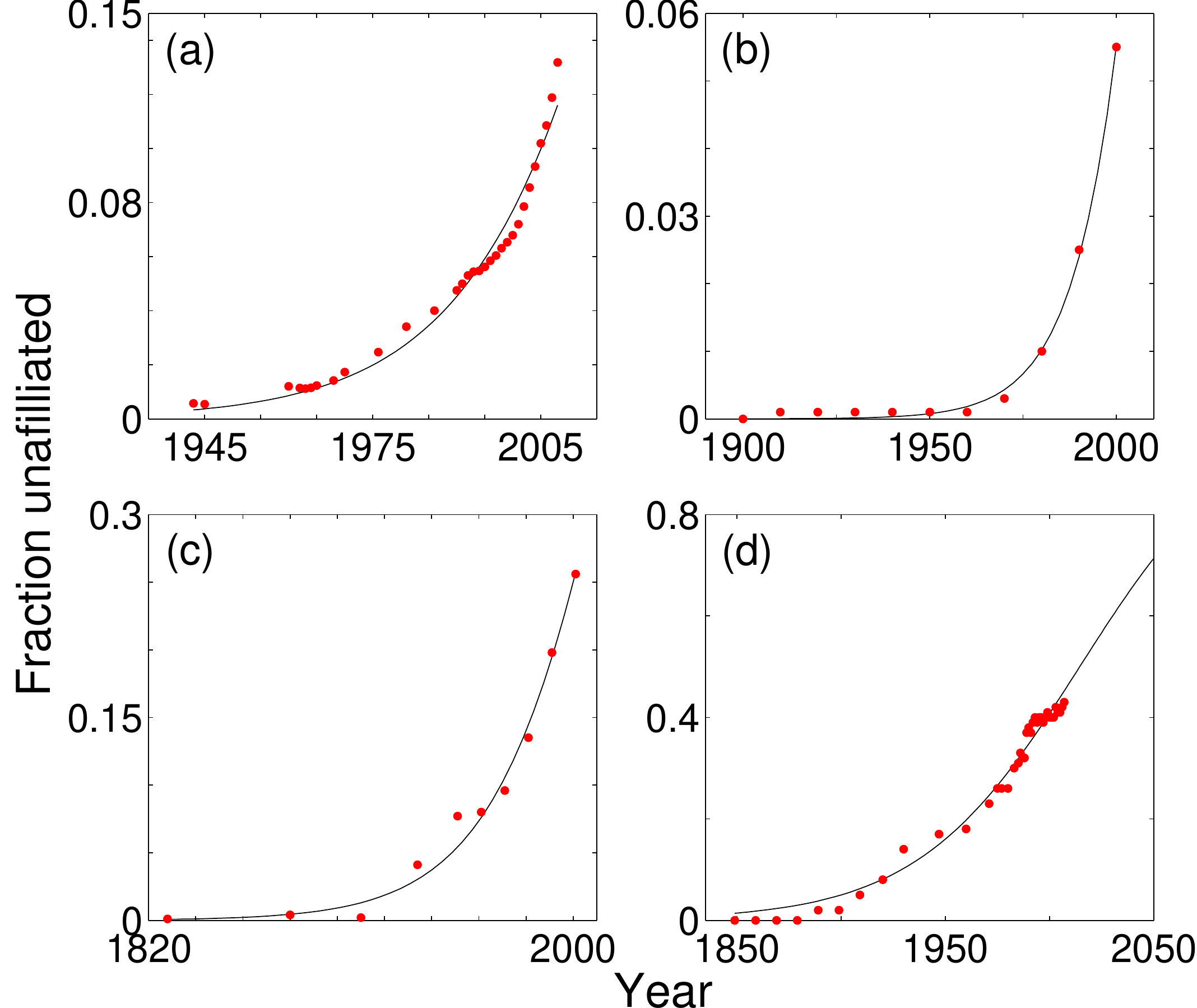}
\caption{Percentage religiously unaffiliated versus time in four regions: (a) the autonomous Aland islands region of Finland, (b) Schwyz Canton in Switzerland, (c) Vienna Province in Austria, (d) the Netherlands.  Red dots indicate data points from census surveys, black lines indicate model fits.  Relative utilities for the religiously unaffiliated populations as determined by model fits were $u_x = 0.63, 0.70, 0.58, 0.56$.\label{fig:x_vs_t}}
\end{figure}

In Figure \ref{fig:x_vs_t} we fit the model to historical census data from regions of Finland, Switzerland, Austria, and the Netherlands, four of 85 worldwide locations for which we compiled and analyzed data.  The initial fraction unaffiliated $x_0$ and the perceived utility $u_x$ were varied to optimize the fit to each data set, while $c$ and $a$ were taken to be global.  A broad minimum in the error near $a = 1$ indicated that as a reasonable choice (see Supplementary Material Section \suppsecdata).  Figure \ref{fig:x_vs_t}(d) shows that, if the model is accurate, nearly 70\% of the Netherlands will be non-affiliated by midcentury.

The behavior of the model can be understood analytically for $a = 1$, in which case we have $dx/dt$ = $c x (1 - x)(2 u_x - 1)$: logistic growth.  An analysis of the fixed points of this equation tells us that religion will disappear if its perceived utility is less than that of non-affiliation, regardless of how large a fraction initially adheres to a religion.  However, if $a$ is less than but close to one, a small social group can indefinitely coexist with a large social group.  Even if $a \ge 1$ it is possible that society will reach such a state if model assumptions break down when the population is nearly all one group.

\begin{figure}
\includegraphics[width=8 cm,clip]{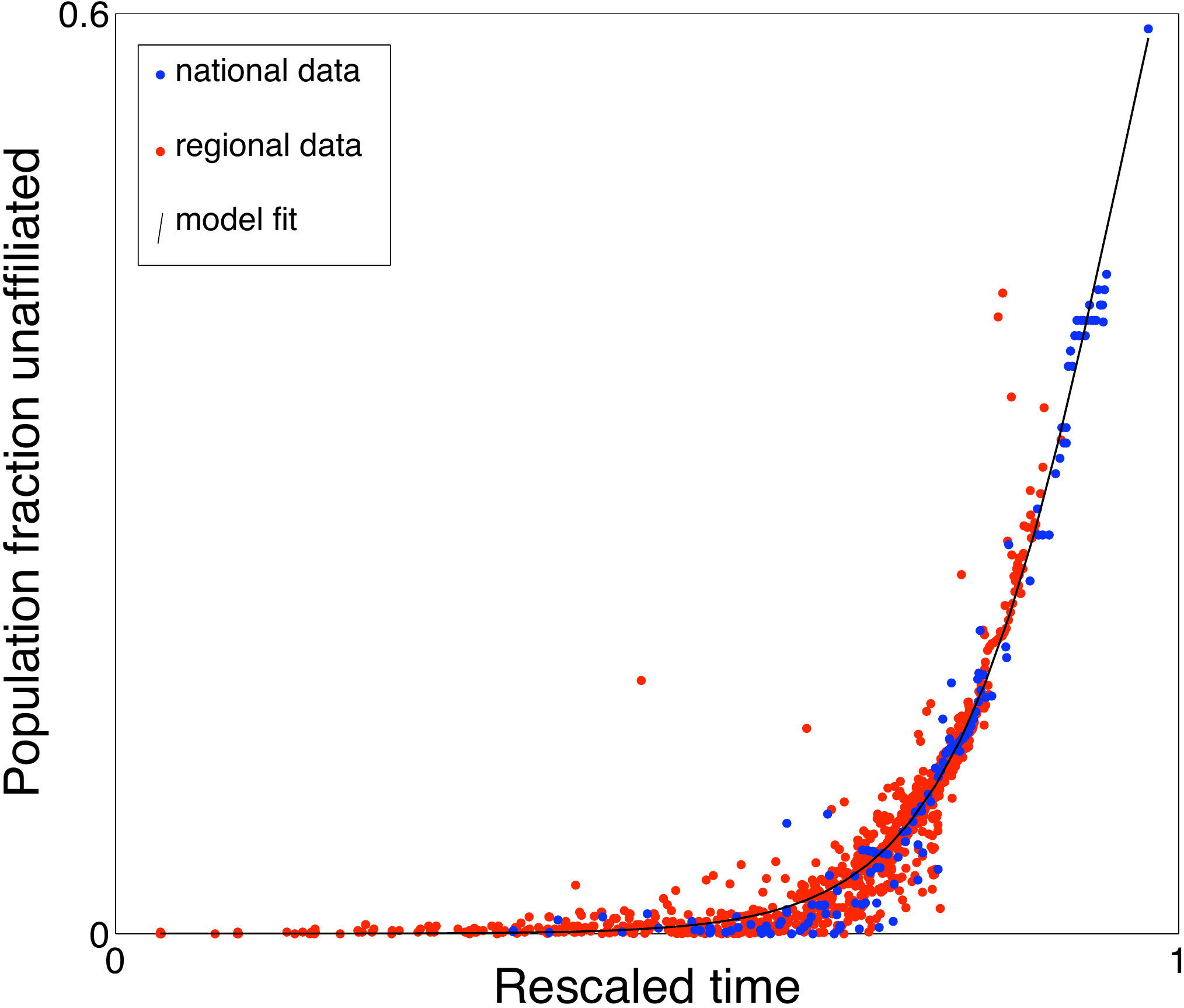}
\caption{All data on changes in religious affiliation with time (85 data sets).  Time has been rescaled so data sets lie on top of one another and the solution curve with $u_x=0.65$.  Red dots correspond to regions within countries, while blue dots correspond to entire countries. Black line indicates model prediction for $u_x=0.65$.\label{fig:alldata}}
\end{figure}

Figure \ref{fig:alldata} shows the totality of the data collected and a comparison to the prediction of Eq.~\eqref{eq1} with $a=1$, demonstrating the general agreement with our model.  Time has been rescaled in each data set and the origin shifted so that they lie on top of one another.  See Supplementary Material Section \suppsecstab~for more details.

Our assumption that the perceived utility of a social group remains constant may be approximately true for long stretches of time, but there may also be abrupt changes in perceived utility, a possibility that is not included in the model.  We speculate that for most of human history, the perceived utility of religion was high and of non-affiliation low.  Religiously non-affiliated people persisted but in small numbers.  With the birth of modern secular societies, the perceived utility of adherence to religion versus non-affiliation has changed significantly in numerous countries\cite{zuckerman07}, such as those with census data shown in Fig.~\ref{fig:x_vs_t}, and the United States, where non-affiliation is growing rapidly\cite{pew09}.

One might ask whether our model explains data better than a simple empirical curve.  Logistic growth would be a reasonable null hypothesis for the observed data, but here we have provided a theoretical framework for expecting a more general growth law \eqref{eq1}, and have shown that data suggest logistic growth as a particular case of the general law.  Our framework includes a rational mathematical foundation for the observed growth law.

\section{Generalizations}

We have thus far assumed that society is highly interconnected in the sense that individual benefits stem from membership in the group that has an overall majority.  For that reason, the model as written is best applied on a small spatial scale where interaction is more nearly all-to-all.  We can generalize this model to include the effects of social networks:  rather than an individual deriving benefits from membership in the global majority group, he or she will instead benefit from belonging to the local majority among his or her social contacts\cite{krapivsky03, galam97}.  In order to define ``local,'' however, we must introduce either a spatial dimension to the problem, or a network defining social interaction.  On a network, Eq.~\eqref{eq1} becomes:
\begin{equation} \label{eq:net}
	\frac{d \langle R_i \rangle}{dt} = (1-\langle R_i \rangle )P_{yx}(x_i,u_x) -  \\\langle R_i \rangle  P_{yx}(1-x_i, 1-u_x)
\end{equation}
where
\begin{equation} \label{eq:localx}
	x_{i} = \left. \sum_{j=1}^N \boldA_{ij} R_j \middle/ \sum_{j=1}^N \boldA_{ij}  \right.
\end{equation}
defines the local mean religious affiliation, $\boldA$ is a binary adjacency matrix defining the social network and $R$ is a binary religious affiliation vector (1 indicates membership in group X).  An ensemble average has been assumed in order to write a derivative for the expected religious affiliation $\left<R_i(t)\right>$ in \eqref{eq:net}, since this system is stochastic rather than deterministic. In the all-to-all coupling limit, $\boldA=\mathbf{1}$ and $x_i=\xbar$, so \eqref{eq:net} reduces to \eqref{eq1}.


A further generalization to a continuous system with arbitrary coupling can be constructed with the introduction of a spatial dimension.  The spatial coordinate $\xi$ will be allowed to vary from $-1$ to $1$ with a normalized coupling kernel $G(\xi,\xi')$ determining the strength of connection between spatial coordinates $\xi$ and $\xi'$.  The religious affiliation variable $R$ now varies spatially and temporally with $0 \le R(\xi,t) \le 1$, so individuals may have varying degrees of affiliation.  Then the dynamics of $R$ satisfy
\begin{equation} \label{dRdt}
	\frac{\partial R}{\partial t} = (1-R)P_{yx}(x,u_x) - R P_{yx}(1-x,1-u_x)
\end{equation}
in analogy with the discrete system.  Here $x$ represents the local mean religious affiliation,
\begin{equation} \label{xdef}
	x(\xi, t) = \int_{-1}^{1} G(\xi,\xi') R(\xi', t) d\xi'~.
\end{equation}
Note that simulation of Eq.~\eqref{eq:net} with continuous real-valued $R$ and large $N$ is equivalent to integration of Eq.~\eqref{dRdt} with appropriate initial conditions and appropriately chosen $G(\xi,\xi')$. This is because \eqref{eq:net} goes from a stochastic system for binary $R$, to a deterministic system for real $R \in [0,1]$.

In the case of all-to-all coupling, $G(\xi,\xi')=1/2$,  and $x(\xi,t) = \frac{1}{2} \int_{-1}^{1} R(\xi',t) d\xi' = \Rbar(t)$, independent of space, where $\Rbar$ is the spatially averaged value of $R$.
Then \eqref{dRdt} becomes
\begin{equation} \label{dRdtall}
	\frac{\partial R}{\partial t} = (1-R)P_{yx}(\Rbar,u_x) - R P_{yx}(1-\Rbar,1-u_x)~.
\end{equation}

If at some time $t$ $R(\xi,t) = R_0(t)$ is independent of space, then $\Rbar(t)=R_0(t)$ and Eq.~\eqref{dRdtall} becomes
\begin{equation} \label{dRdtallconstR}
	\frac{\partial R_0}{\partial t} = (1-R_0)P_{yx}(R_0,u_x) - R_0 P_{yx}(1-R_0,1-u_x)~,
\end{equation}
which follows dynamics identical to the original two-group discrete system \eqref{eq1}.

We can impose perturbations to both the coupling kernel (i.e., the social network structure) and the spatial distribution of $R$ values to examine the stability of this system and the robustness of our results for the all-to-all case.  One very destabilizing perturbation consists of perturbing the system towards two separate clusters with different $R$ values.  These clusters might represent a polarized society that consists of two social cliques in which members of each clique are more strongly connected to members of their clique than to members of the other clique.  Mathematically, this can be written as $G(\xi,\xi') = \half + \half \delta (2 H(\xi)-1) (2 H(\xi')-1)$, where $\delta$ is a small parameter ($\delta \ll 1$) that determines the amplitude of the perturbation and $H(\xi)$ represents the Heaviside step function.  This kernel implies that individuals with the same sign of $\xi$ are more strongly coupled to one-another than they are to individuals with opposite-signed $\xi$.

The above perturbation alone is not sufficient to change the dynamics of the system---a uniform state $R(\xi,t_0) = R_0$ will still evolve according to the dynamics of the original system \eqref{eq1} (See Supplementary Material Section \suppsecpert).

We add a further perturbation to the spatial distribution by imposing $R(\xi,t_0) = R_0 + \epsilon \sign(\xi)$, where $\epsilon$ is a small parameter.  This should conspire with the perturbed coupling kernel to maximally destabilize the uniform state.

Surprisingly, an analysis of the resulting dynamics reveals that this perturbed system must ultimately tend to the same steady state as the original unperturbed system.  Furthermore, the spatial perturbation must eventually decay exponentially, although an initial growth is possible.
(See Supplementary Material Section \suppsecpert~for more details on the perturbative analysis.)
  
The implication of this analysis is that systems that are nearly all-to-all should behave very similarly to an all-to-all system.  In the next Section we describe a numerical experiment that tests this prediction.

\section{Numerical Experiment} \label{sec:numerics}

We design our experiment with the goal of controlling the perturbation from an all-to-all network through a single parameter.  We construct a social network consisting of two all-to-all clusters initially disconnected from one-another, and then add links between any two nodes in opposite clusters with probability $p$.  Thus $p=1$ corresponds to an all-to-all network that should simulate \eqref{eq1}, while $p=0$ leaves the network with two disconnected components.  Small perturbations from all-to-all correspond to $p$ near 1, and $p$ can be related to the coupling kernel perturbation parameter $\delta$ described above as $p = (1-\delta)/(1+\delta)$ (assuming all links in the network have equal weight).  The size of each cluster is determined by the initial condition $x_0$ as $N_X = x_0 N$, $N_Y = (1-x_0)N$.

\begin{figure}
\includegraphics[width=8.6cm,clip]{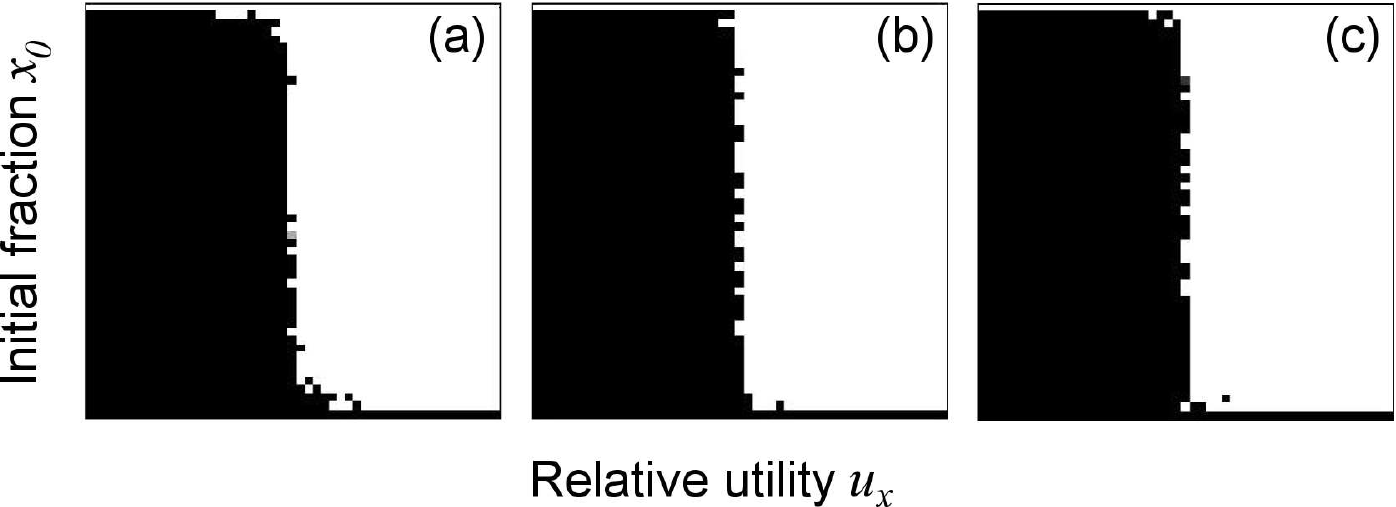}
\caption{Results of simulation of the discrete stochastic model \eqref{eq:net} on a network with two initial clusters weakly coupled to one another. The ratio $p$ of out-group coupling strength to in-group coupling strength is (a) $p=0.01$; (b) $p=0.40$; (c) $p=0.80$ ($N=500$).  Steady states are nearly identical to the predictions of the all-to-all model \eqref{eq1}.\label{fig:discrete_numerics}}
\end{figure}

Figure \ref{fig:discrete_numerics} compares the results of simulation of system \eqref{eq:net} with varying perturbations off of all-to-all.  The theoretical (all-to-all) separatrix between basins of attraction is a vertical line at $u_x=1/2$.  Even when $p=0.01$, when in-group connections are 100 times more numerous than out-group connections, the steady states of the system and basins of attraction remain essentially unchanged.

In the case of the continuous deterministic system \eqref{dRdt}, the equivalent figure to \ref{fig:discrete_numerics} is extremely boring: numerically, the steady states of the perturbed system are indistinguishable from those of the unperturbed all-to-all system, regardless of the value of $p$ (see Supplementary Section \suppsecnumerics~and Figure \suppfignumerics). 

\begin{figure}
\includegraphics[width=8 cm,clip]{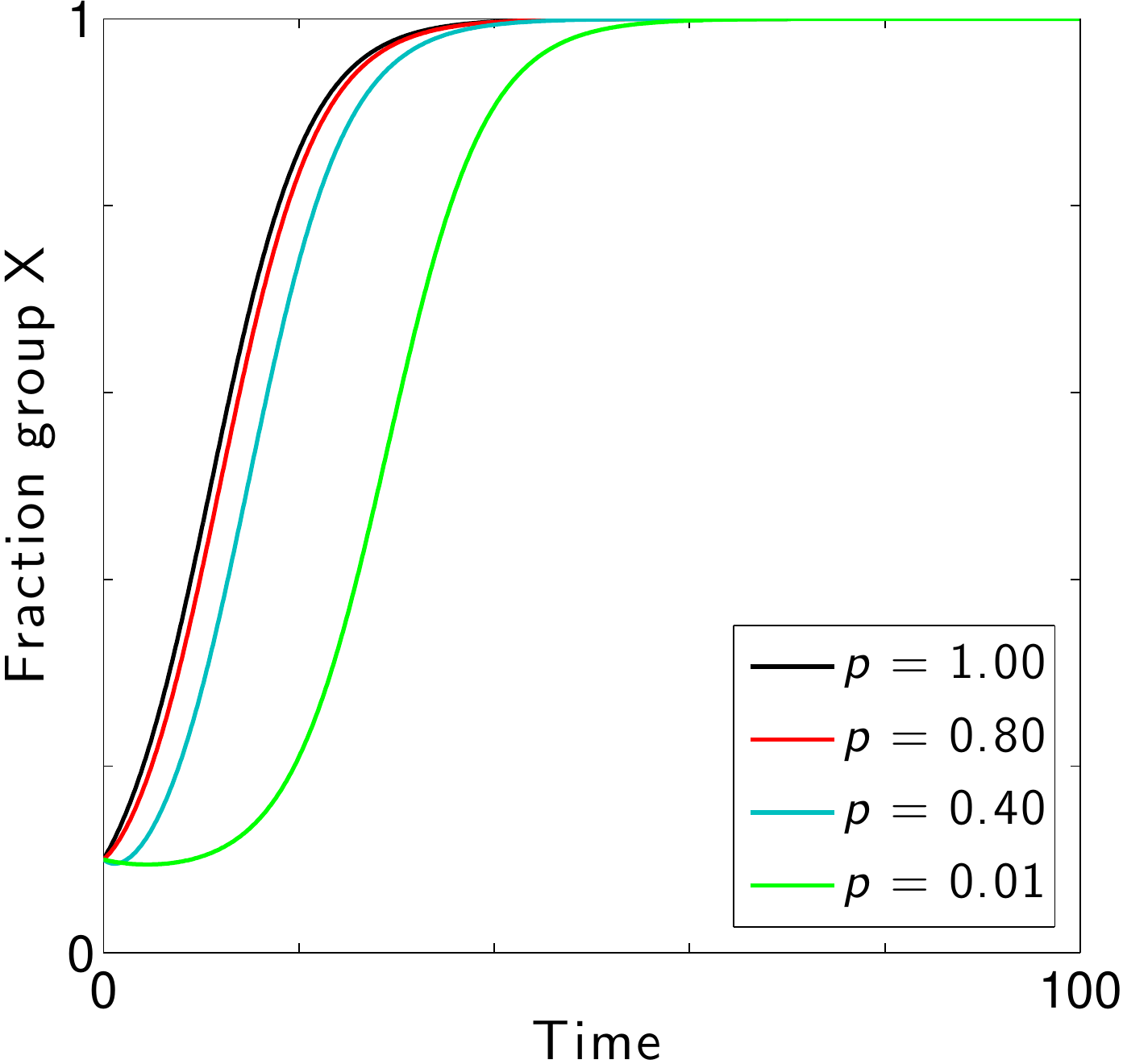}
\caption{Variation in the behavior of system \eqref{dRdt} with increasing perturbation off of all-to-all ($N=500$, $x_0=0.1$, $u=0.6$).  Equivalent values of the perturbation parameter $\delta$ in order of decreasing $p$ are $\delta=0$, $\delta=0.14$, $\delta=0.60$, and $\delta=0.98$.\label{fig:time_delay}}
\end{figure}

The only notable difference between the dynamics of the continuous networked system and the dynamics of the original all-to-all system \eqref{eq1} is a time delay $d$ apparent before the onset of significant shift between groups (see Figure \ref{fig:time_delay}).  We were able to find an approximate expression for that time delay as $d \propto -\ln p / (2u-1)$ (see Supplementary Material Section \suppsecdelay, figures \suppfigxvst~and \suppfigdelay). 

What we have shown by the generalization of the model to include network structure is surprising: even if conformity to a local majority influences group membership, the existence of some out-group connections is enough to drive one group to dominance and the other to extinction.  In the language of references \cite{krapivsky03, benczik08, holme06}, the population will reach the same consensus, despite the existence of individual cliques, as it would without cliques, with only the addition of a time delay.

In a modern secular society there are many opportunities for out-group connections to form due to the prevalence of socially integrated institutions---schools, workplaces, recreational clubs, etc.  Our analysis shows that just a few out-group connections are sufficient to explain the good fit of Eq.~\eqref{eq1} to data, even though Eq.~\eqref{eq1} implicitly assumes all-to-all coupling.

\section{Conclusions}

We have developed a general framework for modeling competition between social groups and analyzed the behavior of the model under modest assumptions. We found that a particular case of the solution fits census data on competition between religious and irreligious segments of modern secular societies in 85 regions around the world. The model indicates that in these societies the perceived utility of religious non-affiliation is greater than that of adhering to a religion, and therefore predicts continued growth of non-affiliation, tending toward the disappearance of religion.  According to our calculations, the steady-state predictions should remain valid under small perturbations to the all-to-all network structure that the model assumes, and, in fact, the all-to-all analysis remains applicable to networks very different from all-to-all. Even an idealized highly polarized society with a two-clique network structure follows the dynamics of our all-to-all model closely, albeit with the introduction of a time delay.  This perturbation analysis suggests why the simple all-to-all model fits data from societies that undoubtedly have more complex network structures.
 
For decades, authors have commented on the surprisingly rapid decline of organized religion in many regions of the world. The work we have presented does not exclude previous models, but provides a new framework for the understanding of different models of human behavior in majority/minority social systems in which groups compete for members. We believe that, with the application of techniques from the mathematics of dynamical systems and perturbation theory, we have gained a deeper understanding of how various assumptions about human behavior will play out in the real world.
\begin{acknowledgments}
This work was funded by Northwestern University and The James S.~McDonnell Foundation.  The authors thank P. Zuckerman for useful correspondence.
\end{acknowledgments}

%

\newpage \qquad \newpage
\includepdf[pages=1]{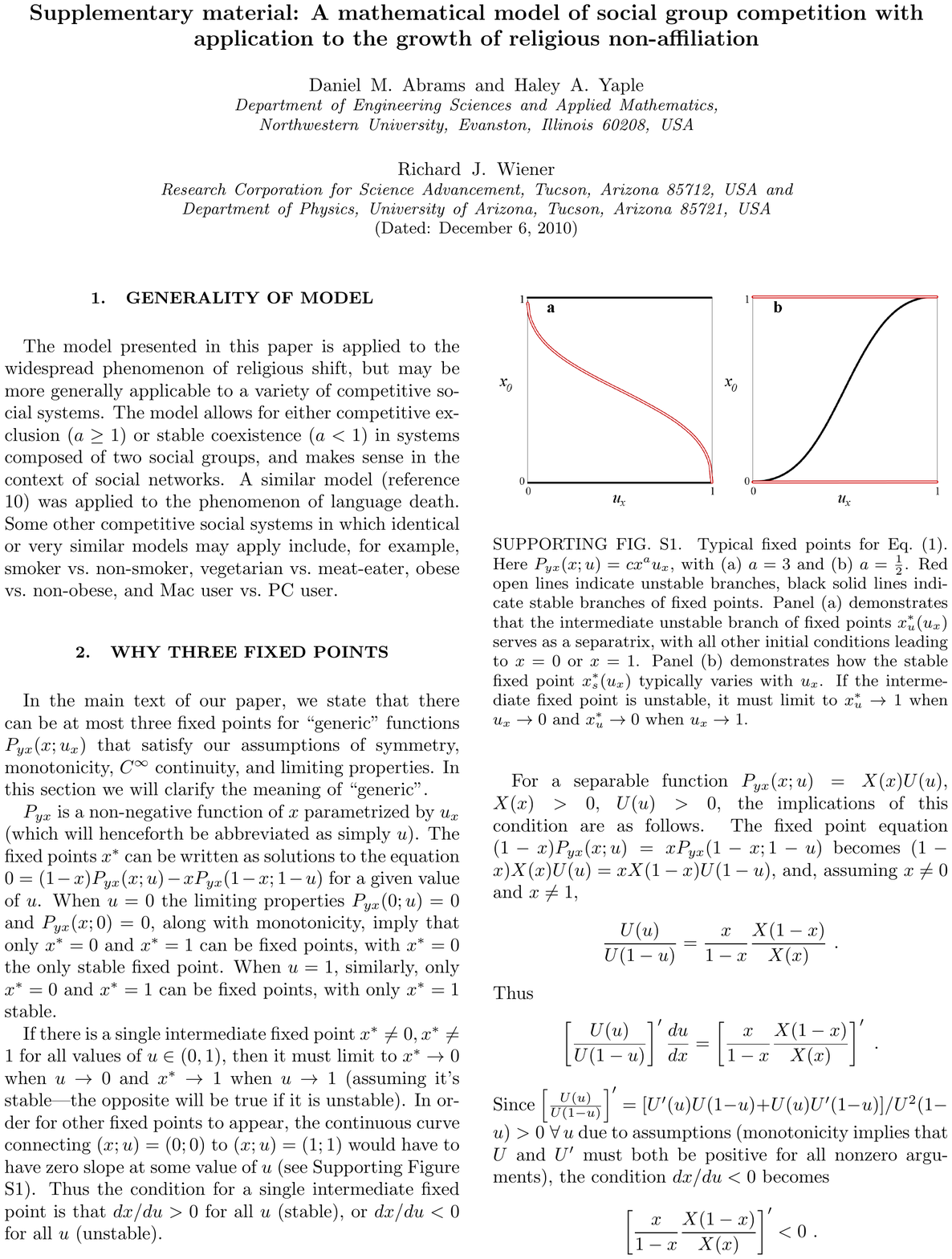}
\newpage \qquad \newpage
\includepdf[pages=2]{Abrams_SI.pdf}  
\newpage \qquad \newpage
\includepdf[pages=3]{Abrams_SI.pdf}
\newpage \qquad \newpage
\includepdf[pages=4]{Abrams_SI.pdf}  
\newpage \qquad \newpage
\includepdf[pages=5]{Abrams_SI.pdf}
\newpage \qquad \newpage


\section*{Supplementary material}

# Supplementary material: A mathematical model of social group competition with application to the growth of religious non-affiliation

Daniel M. Abrams and Haley A. Yaple
*Department of Engineering Sciences and Applied Mathematics,*
*Northwestern University, Evanston, Illinois 60208, USA*

Richard J. Wiener
*Research Corporation for Science Advancement, Tucson, Arizona 85712, USA and*
*Department of Physics, University of Arizona, Tucson, Arizona 85721, USA*
(Dated: December 6, 2010)

## 1. GENERALITY OF MODEL

The model presented in this paper is applied to the widespread phenomenon of religious shift, but may be more generally applicable to a variety of competitive social systems. The model allows for either competitive exclusion ($a \geq 1$) or stable coexistence ($a < 1$) in systems composed of two social groups, and makes sense in the context of social networks. A similar model (reference 10) was applied to the phenomenon of language death. Some other competitive social systems in which identical or very similar models may apply include, for example, smoker vs. non-smoker, vegetarian vs. meat-eater, obese vs. non-obese, and Mac user vs. PC user.

## 2. WHY THREE FIXED POINTS

In the main text of our paper, we state that there can be at most three fixed points for "generic" functions $P_{yx}(x; u_x)$ that satisfy our assumptions of symmetry, monotonicity, $C^\infty$ continuity, and limiting properties. In this section we will clarify the meaning of "generic".

$P_{yx}$ is a non-negative function of $x$ parametrized by $u_x$ (which will henceforth be abbreviated as simply $u$). The fixed points $x^*$ can be written as solutions to the equation $0 = (1-x)P_{yx}(x; u) - xP_{yx}(1-x; 1-u)$ for a given value of $u$. When $u = 0$ the limiting properties $P_{yx}(0; u) = 0$ and $P_{yx}(x; 0) = 0$, along with monotonicity, imply that only $x^* = 0$ and $x^* = 1$ can be fixed points, with $x^* = 0$ the only stable fixed point. When $u = 1$, similarly, only $x^* = 0$ and $x^* = 1$ can be fixed points, with only $x^* = 1$ stable.

If there is a single intermediate fixed point $x^* \neq 0, x^* \neq 1$ for all values of $u \in (0, 1)$, then it must limit to $x^* \to 0$ when $u \to 0$ and $x^* \to 1$ when $u \to 1$ (assuming it's stable—the opposite will be true if it is unstable). In order for other fixed points to appear, the continuous curve connecting $(x; u) = (0; 0)$ to $(x; u) = (1; 1)$ would have to have zero slope at some value of $u$ (see Supporting Figure S1). Thus the condition for a single intermediate fixed point is that $dx/du > 0$ for all $u$ (stable), or $dx/du < 0$ for all $u$ (unstable).

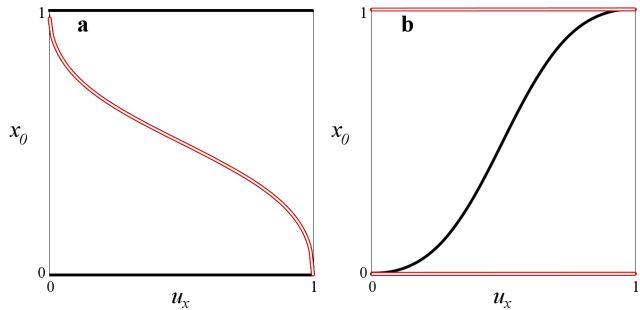

SUPPORTING FIG. S1. Typical fixed points for Eq. (1). Here $P_{yx}(x; u) = cx^a u_x$, with (a) $a = 3$ and (b) $a = \frac{1}{2}$. Red open lines indicate unstable branches, black solid lines indicate stable branches of fixed points. Panel (a) demonstrates that the intermediate unstable branch of fixed points $x_u^*(u_x)$ serves as a separatrix, with all other initial conditions leading to $x = 0$ or $x = 1$. Panel (b) demonstrates how the stable fixed point $x_s^*(u_x)$ typically varies with $u_x$. If the intermediate fixed point is unstable, it must limit to $x_u^* \to 1$ when $u_x \to 0$ and $x_u^* \to 0$ when $u_x \to 1$.

For a separable function $P_{yx}(x; u) = X(x)U(u)$, $X(x) > 0$, $U(u) > 0$, the implications of this condition are as follows. The fixed point equation $(1 - x)P_{yx}(x; u) = xP_{yx}(1 - x; 1 - u)$ becomes $(1 - x)X(x)U(u) = xX(1 - x)U(1 - u)$, and, assuming $x \neq 0$ and $x \neq 1$,

$$\frac{U(u)}{U(1-u)} = \frac{x}{1-x}\frac{X(1-x)}{X(x)} \ .$$

Thus

$$\left[\frac{U(u)}{U(1-u)}\right]' \frac{du}{dx} = \left[\frac{x}{1-x}\frac{X(1-x)}{X(x)}\right]' \ .$$

Since $\left[\frac{U(u)}{U(1-u)}\right]' = [U'(u)U(1-u) + U(u)U'(1-u)]/U^2(1-u) > 0 \ \forall u$ due to assumptions (monotonicity implies that $U$ and $U'$ must both be positive for all nonzero arguments), the condition $dx/du < 0$ becomes

$$\left[\frac{x}{1-x}\frac{X(1-x)}{X(x)}\right]' < 0 \ .$$

This can be simplified to

$$\frac{X'(x)}{X(x)} + \frac{X'(1-x)}{X(1-x)} > \frac{1}{x} + \frac{1}{1-x}.$$

The direction of the inequality would be reversed for a stable intermediate fixed point. Note that a *sufficient* condition would be $X'/X > 1/x$. This, or the same condition with the inequality reversed, is clearly satisfied for any power law form $P_{yx}(x;u) \sim x^a$, $a \neq 1$. It is also satisfied for any function with a monotonic first derivative $X'(x)$ (Sketch of proof: Let $X'(x) = X'_0 + f(x)$, where $f(x) \equiv \int_0^x X''(\xi)d\xi$ is a monotonically increasing function. Then $xX'(x) = xX'_0 + xf(x)$ and $X(x) = \int_0^x X'(\xi)d\xi = xX'_0 + \int_0^x f(\xi)d\xi$. Thus $xX'(x) - X(x) = xf(x) - \int_0^x f(\xi)d\xi$. That last quantity is necessarily greater than zero for any monotonically increasing $f(x)$, and therefore $xX'(x) > X(x)$, or $X'/X > 1/x$.)

An analogous result holds for inseparable functions $P_{yx}(x;u)$. Using the same approach, the condition is:

$$\frac{1}{x}P'_{yx}(x;u) + \frac{1}{1-x}P'_{yx}(1-x;1-u) > \frac{1}{x^2}P_{yx}(x;u) + \frac{1}{(1-x)^2}P_{yx}(1-x;1-u),$$

where prime notation represents a derivative with respect to the argument, not the parameter. Thus a sufficient condition is $P'_{yx}(x;u)/P_{yx}(x;u) > 1/x$ for all $u$. The full condition is satisfied by any function for which curvature doesn't change sign.

## 3. STABILITY OF FIXED POINTS

Examine the stability of the fixed point at $x = 0$ (and note that the same argument will work for the stability of the fixed point at $x = 1$). Set $x = \eta$, a small perturbation from $x = 0$. Then

$$\dot{\eta} = (1-\eta)P_{yx}(\eta;u) - \eta P_{yx}(1-\eta;1-u)$$
$$\approx P_{yx}(0;u) + \eta\left[P'_{yx}(0;u) - P_{yx}(0;u) - P_{yx}(1;1-u)\right]$$
$$= -\eta\left[P_{yx}(1;1-u) - P'_{yx}(0;u)\right]$$

to $\mathcal{O}(\eta^2)$. So the fixed point $x^* = 0$ is stable to small perturbations if $P_{yx}(1;1-u) > P'_{yx}(0;u)$. For a power law $P_{yx} \sim x^a$, this will be true only when $a > 1$. The fixed point $x^* = 0$ will be unstable for $a < 0$, and its stability will depend on the sign of $u - \frac{1}{2}$ when $a = 1$.

The stability of the intermediate fixed point is fully determined once the stabilities of the two endpoints $x^* = 0$ and $x^* = 1$ are known. Because it is a one-dimensional flow, the intermediate fixed point must be stable when the endpoints are unstable, and vice-versa when the endpoints are stable (see Figure S2).

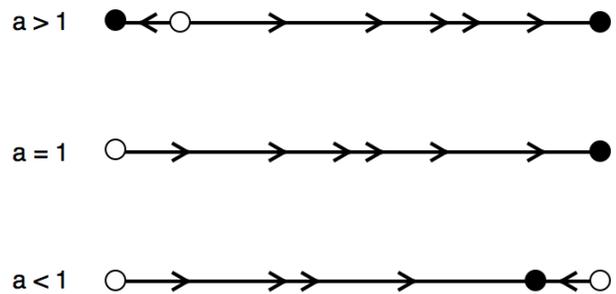

SUPPORTING FIG. S2. The flow in $x$ for various values of the constant $a$. Filled circles indicate stable fixed points, while open circles indicate unstable fixed points. The leftmost fixed points correspond to $x = 0$, while the rightmost fixed points correspond to $x = 1$.

## 4. DATA SETS AND MODEL FITS

Data used in validating this model originated in census surveys from a range of countries worldwide. A total of 85 data sets had 5 or more independent data points. These came from various regions of 9 different countries: Australia, Austria, Canada, the Czech Republic, Finland, Ireland, the Netherlands, New Zealand, and Switzerland.

Fitting was done by minimizing root-mean-square errors. Using a functional form $P_{yx} = cx^a u$, the parameters $c$ and $a$ were taken to be universal while the parameter $u$ was allowed to vary with each data set. This was accomplished by simultaneously optimizing $c$, $a$, and $u_1...u_N$ such that the RMS error summed over all $N$ data sets was minimized.

Supporting Figure S3 shows how that summed error varied with the parameter $a$. We chose $a \approx 1$ for the fits discussed in this paper both for simplicity and because of the broad minimum visible around $a = 1$. The parameter $c$, which simply sets a time scale, was approximately 0.2.

## 5. PERTURBATION OF NETWORK STRUCTURE

In this section we examine in greater depth the implications of Eq. (4), a continuous deterministic system with arbitrary coupling.

### All-to-all coupling

If $G(\xi,\xi') = 1/2$ then there is uniform all-to-all coupling, and we see that $x(\xi,t) = \frac{1}{2}\int_{-1}^1 R(\xi',t)d\xi' = \bar{R}(t)$, independent of space, where $\bar{R}$ is the spatially averaged value of $R$.

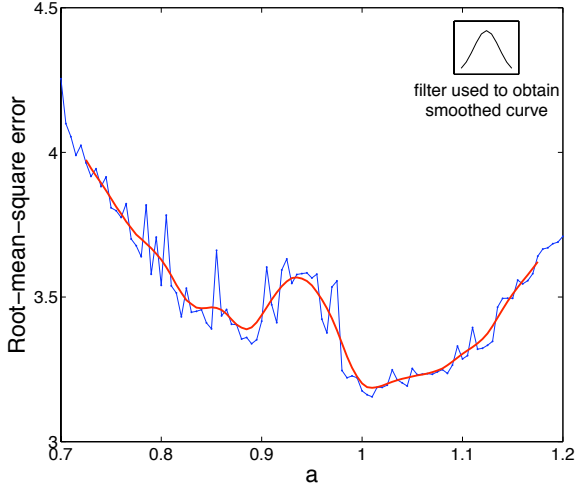

SUPPORTING FIG. S3. Summed root-mean-square error over all data sets versus parameter $a$ in $P_{yx} = cx^a u_x$. The error was calculated by finding the combination of parameters $c$, $x_{0i}$ and $u_{xi}$ (where $i$ varies over all data sets) that minimized the root mean square error between the model predictions and the data. Blue curve indicates exact error calculations, red indicates smoothed error after convolution with a Gaussian (inset). Note that there appears to be a broad minimum near $a = 1$.

Then (4) becomes

$$\frac{\partial R}{\partial t} = (1-R)P_{yx}(\bar{R}; u) - R P_{yx}(1-\bar{R}; 1-u) \quad (S1)$$

If at some time $t^*$ $R(\xi, t^*) = R_0(t^*)$ is independent of space, then $\bar{R}(t^*) = R_0(t^*)$ and Eq. (S1) becomes

$$\frac{\partial R_0}{\partial t} = (1-R_0)P_{yx}(R_0; u) - R_0 P_{yx}(1-R_0; 1-u) \text{ , (S2)}$$

which follows dynamics identical to the original two-group discrete system.

**Perturbation off of uniform $R$ with all-to-all coupling**

We impose a destabilizing perturbation such that the portion of the population with $\xi < 0$ has lower $R$ and the portion with $\xi > 0$ has higher $R$, i.e.,

$$R(\xi, t_0) = R_0 + \epsilon \, \text{sgn}(\xi) \text{ , (S3)}$$

where $\epsilon$ is a small parameter. Then $x(\xi) = \frac{1}{2}\int_{-1}^{1} R(\xi')d\xi' = R_0$, and from (4) we get

$$\frac{\partial R}{\partial t} = (1 - R_0 - \epsilon \, \text{sgn}(\xi))P_{yx}(R_0; u) \\ - (R_0 + \epsilon \, \text{sgn}(\xi))P_{yx}(1-R_0; 1-u) \quad (S4)$$

We can also look at the dynamics of the mean religious affiliation $\bar{R}$,

$$\frac{\partial \bar{R}}{\partial t} = \frac{\partial}{\partial t}\left(\frac{1}{2}\int_{-1}^{1} R(\xi', t)d\xi'\right) = \frac{1}{2}\int_{-1}^{1} \frac{\partial R(\xi', t)}{\partial t}d\xi' \text{ .} \quad (S5)$$

Plugging (S4) into (S5) and simplifying gives

$$\frac{\partial \bar{R}}{\partial t} = \frac{\partial R_0}{\partial t} = (1-R_0)P_{yx}(R_0; u) - R_0 P_{yx}(1-R_0; 1-u) \text{ ,} \quad (S6)$$

so the mean religious affiliation $\bar{R}(t)$ continues to follow the dynamics of the original system despite the perturbation.

Rearranging (S4), we see

$$\frac{\partial R}{\partial t} = \frac{\partial R_0}{\partial t} - \epsilon \, \text{sgn}(\xi)(P_{yx}(\bar{R}; u) + P_{yx}(1-\bar{R}; 1-u)) \text{ ,} \quad (S7)$$

and direct differentiation of (S3) yields

$$\frac{\partial R}{\partial t} = \frac{\partial R_0}{\partial t} + \frac{\partial \epsilon}{\partial t} \, \text{sgn}(\xi) \text{ .} \quad (S8)$$

Equating these expressions yields a differential equation for $\epsilon(t)$:

$$\frac{\partial \epsilon}{\partial t} = -\epsilon(P_{yx}(\bar{R}; u) + P_{yx}(1-\bar{R}; 1-u)) \text{ .} \quad (S9)$$

Note that $\epsilon$ remains independent of the spatial coordinate, and that $\epsilon \to 0$ as $t \to \infty$, for any initial condition (the time constant may vary with the parameter $u$ and the state $\bar{R}$). So the initial perturbation must damp out, and the system must evolve to a single affiliation as $t \to \infty$, just as the original system (1) did.

**Non-uniform coupling**

We consider the case of non-uniform spatial coupling as the continuum limit of a discrete network where the links are nearly but not quite all-to-all. In that case, a very destabilizing perturbation would be one in which the network is segregated into two clusters, each one more strongly coupled internally than externally. As described in the main text, one kernel representing such coupling is

$$G(\xi, \xi') = \frac{1}{2} + \frac{1}{2}\delta(2H(\xi) - 1)(2H(\xi') - 1) \text{ ,} \quad (S10)$$

where $\delta$ is a small parameter ($\delta \ll 1$) that determines the amplitude of the perturbation and $H(\xi)$ represents the Heaviside step function.

If the initial state of the population is uniform such that $R(\xi, t_0) = R_0$ then $x(\xi, t_0) = \int_{-1}^{1} G(\xi, \xi')R_0 d\xi' = R_0$ and $R$ will satisfy Eq. (4), giving

$$\frac{\partial R_0}{\partial t} = (1-R_0)P_{yx}(R_0; u) - R_0 P_{yx}(1-R_0; 1-u) \text{ .} \quad (S11)$$

Thus $R$ will remain uniform in space and will follow the same dynamics as the original system, *despite* a non-uniform coupling kernel of arbitrary amplitude.

### Perturbation off of uniform $R$ with non-uniform coupling

As before, we impose a destabilizing perturbation such that the portion of the population with $\xi < 0$ has lower $R$ and the portion with $\xi > 0$ has higher $R$, i.e., $R(\xi, t_0) = R_0 + \epsilon \operatorname{sgn}(\xi)$, where $\epsilon$ is again a small parameter. This should conspire with the perturbed coupling kernel to maximally destabilize the uniform state.

Now (5) gives

$$x(\xi, t) = (R_0 - \epsilon) \int_{-1}^{0} G(\xi, \xi') d\xi' + (R_0 + \epsilon) \int_{0}^{1} G(\xi, \xi') d\xi'$$
$$= R_0 + \epsilon \delta \operatorname{sgn}(\xi), \qquad (S12)$$

and from (4) we get

$$\frac{\partial R}{\partial t} = (1 - R_0 - \epsilon \operatorname{sgn}(\xi)) P_{yx}(R_0 + \epsilon \delta \operatorname{sgn}(\xi); u)$$
$$- (R_0 + \epsilon \operatorname{sgn}(\xi)) P_{yx}(1 - R_0 - \epsilon \delta \operatorname{sgn}(\xi); 1 - u). \qquad (S13)$$

Directly differentiating $R(\xi, t_0) = R_0 + \epsilon \operatorname{sgn}(\xi)$, then rearranging terms on the right-hand-side of (S13) gives

$$\frac{\partial R_0}{\partial t} + \operatorname{sgn}(\xi) \frac{\partial \epsilon}{\partial t} = (1 - R_0) P_{yx}(R_0 + \epsilon \delta \operatorname{sgn}(\xi); u)$$
$$- R_0 P_{yx}(1 - R_0 - \epsilon \delta \operatorname{sgn}(\xi); 1 - u) - \epsilon \operatorname{sgn}(\xi) \left[ P_{yx}(R_0 + \epsilon \delta \operatorname{sgn}(\xi); u) + P_{yx}(1 - R_0 - \epsilon \delta \operatorname{sgn}(\xi); 1 - u) \right]. \qquad (S14)$$

Now calculate $\frac{\partial R_0}{\partial t}$ directly. Note that $R_0 = \bar{R} = \frac{1}{2} \int_{-1}^{1} R(\xi', t) d\xi'$, so

$$\frac{\partial R_0}{\partial t} = \frac{\partial}{\partial t} \left( \frac{1}{2} \int_{-1}^{1} R(\xi', t) d\xi' \right) = \frac{1}{2} \int_{-1}^{1} \frac{\partial R(\xi', t)}{\partial t} d\xi'$$
$$= \frac{1}{2} \int_{-1}^{1} \left[ (1 - R_0 - \epsilon \operatorname{sgn}(\xi')) P_{yx}(R_0 + \epsilon \delta \operatorname{sgn}(\xi'); u) - (R_0 + \epsilon \operatorname{sgn}(\xi')) P_{yx}(1 - R_0 - \epsilon \delta \operatorname{sgn}(\xi'); 1 - u) \right] d\xi'$$
$$= \frac{1}{2}(1 - R_0) \left[ P_{yx}(R_0 - \epsilon \delta; u) + P_{yx}(R_0 + \epsilon \delta; u) \right] - \frac{1}{2} R_0 \left[ P_{yx}(1 - R_0 + \epsilon \delta; 1 - u) + P_{yx}(1 - R_0 - \epsilon \delta; 1 - u) \right]$$
$$+ \frac{1}{2} \epsilon \left[ P_{yx}(R_0 - \epsilon \delta; u) - P_{yx}(R_0 + \epsilon \delta; u) \right] + \frac{1}{2} \epsilon \left[ P_{yx}(1 - R_0 + \epsilon \delta; 1 - u) - P_{yx}(1 - R_0 - \epsilon \delta; 1 - u) \right].$$

Taylor expanding in both $\epsilon$ and $\delta$ eliminates all first order terms in $\epsilon$, leaving

$$\frac{\partial R_0}{\partial t} = (1 - R_0) P_{yx}(R_0; u) - R_0 P_{yx}(1 - R_0; 1 - u) + \mathcal{O}(\epsilon^2 \delta), \qquad (S15)$$

so the assumption that the mean religious affiliation follows the dynamics of the original unperturbed system is well justified.

Similarly Taylor expanding Eq. (S14) to first order in $\epsilon$ and $\delta$ allows canceling of the $\partial R_0 / \partial t$ terms on either side (using (S15)), leaving an equation in $\epsilon$:

$$\frac{\partial \epsilon}{\partial t} = - \epsilon \Big\{ P_{yx}(R_0; u) + P_{yx}(1 - R_0; 1 - u)$$
$$- \delta \big[ (1 - R_0) P'_{yx}(R_0; u)$$
$$+ R_0 P'_{yx}(1 - R_0; 1 - u) \big] \Big\}. \qquad (S16)$$

The sign of the quantity in braces in (S16) determines the stability of the uniform spatial state. It's clear that for sufficiently small $\delta$, the uniform state will *always* be stable. However, in systems with an unstable intermediate fixed point (or no intermediate fixed point), the uniform state will remain stable even when the quantity in braces is initially positive! This is because Eq. (S15) will still hold for small $\epsilon$, making $R_0$ approach a steady state value $R_0^* = 0$ or $R_0^* = 1$ from the original system. Since $\epsilon < \min(R_0, 1 - R_0)$ is required to maintain variables in the allowed domain, $\epsilon$ may initially grow, but it will have to eventually shrink as $R_0 \to R_0^*$.

The above further shows that $\epsilon$ does not develop any additional spatial structure, so an initial state with $R = R_0 + \epsilon \operatorname{sgn}(\xi)$ will maintain such a spatial structure as $R_0$ and $\epsilon$ evolve in time.

This calculation demonstrates that an understanding of the simple all-to-all discrete system gives insight into the more complex problem of religious shift on a social

network. In numerical experiment, the results of the perturbation calculation described here remain valid even for very sparse networks quite different from all-to-all.

## 6. NUMERICS

In the main text, we describe a numerical experiment that we performed on systems (2) and (4). Figure 3 of the main text shows the results of that experiment with a simulated size $N = 500$, and in Figure S4, we show that the all-to-all system (1) becomes a good match to the discrete stochastic system (2) as the number of nodes increases (thus explaining why Figure 3 is well predicted by understanding the all-to-all system).

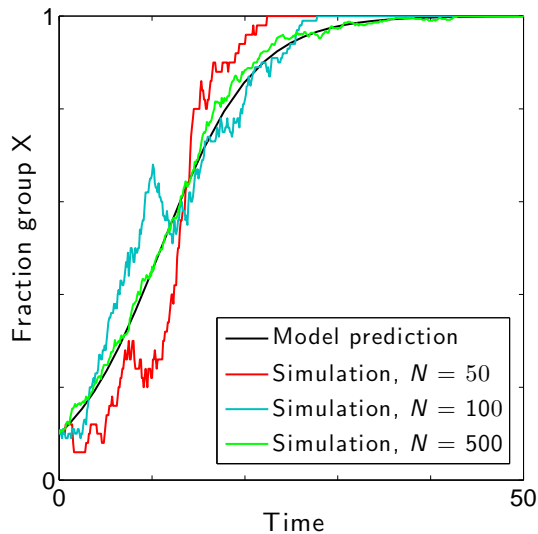

SUPPORTING FIG. S4. Comparison of simulation of discrete stochastic system (2) to model predictions for various system sizes (all-to-all coupling). Here $x_0 = 0.1$ and the total size of the network is 50 (red), 100 (blue), or 500 (green). The solid black line represents the solution to Eq. (1), the large $N$ limit of this model.

We also performed a our numerical experiment on (4), the continuous deterministic generalization of (1). Results are presented in Figure S5, where the steady states are indistinguishable from the all-to-all model.

## 7. TIME DELAY

We define the effective time delay $d$ to be the delay between the perturbed solution (not all-to-all) and the all-to-all solution (the logistic function, when $P_{yx} = cx^a u$ with $a \approx 1$), as measured when $\bar{R}$ has risen halfway to its asymptotic value of 1 (we assume a rising function with no loss of generality: the symmetric case of a decaying function can be examined under the change of variables $u \mapsto 1-u, x_0 \mapsto 1-x_0$). Thus $d$ is the difference in the

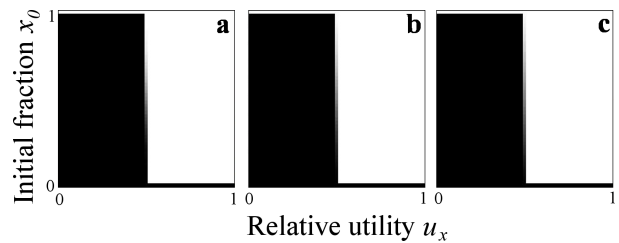

SUPPORTING FIG. S5. Results of simulation of the continuous deterministic system (4) on a network with two initial clusters weakly coupled to one another. The ratio $p$ of out-group coupling strength to in-group coupling strength is (a) $p = 0.01$; (b) $p = 0.40$; (c) $p = 0.80$ ($N = 500$). When $u_x = 1/2$, all points are fixed points, so the initial condition determines the final state. Steady states are indistinguishable from those of the all-to-all model (1) despite the non-uniform coupling and inhomogeneous initial conditions.

time $t_c$ when a solution $\bar{R}(t_c) = \frac{1}{2}(1+R_0)$ and $t_{c_0}$ when $\bar{R}_{all-to-all}(t_{c_0}) = \frac{1}{2}(1+R_0)$. We observe this quantity to increase monotonically with the perturbation off of all-to-all $\delta$—see Figure S6 for typical behavior at various values of $\delta$. In the limit that $p \ll 1$ ($\delta$ near 1, nearly two separate clusters) we find that the curve is well approximated by $d \propto -\ln(p)/(2u-1)$.

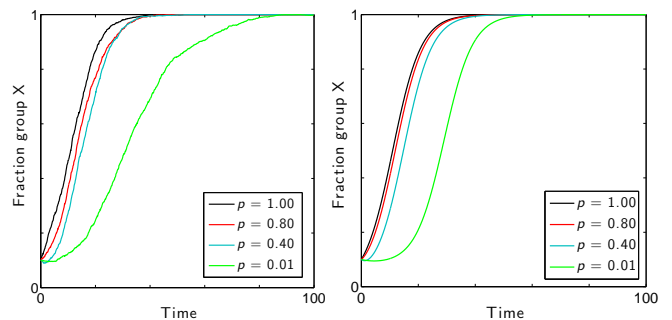

SUPPORTING FIG. S6. Variation in the behavior of systems (2) and (4) with increasing perturbation off of all-to-all. This illustrates delay time as inter-cluster connection probability $p$ varies. Equivalent values of the perturbation parameter $\delta$ in order of decreasing $p$ are $\delta = 0$, $\delta = 0.14$, $\delta = 0.60$, and $\delta = 0.98$. Left panel: Discrete stochastic system (2) (ensemble averages of 10 realizations). Right panel: Continuous deterministic system (4). For all simulations $x(0) = 0.1$, $u = 0.6$ and $N = 500$.

We find this form by assuming $\bar{R}(t) \approx \bar{R}_0 + \eta y(t) + \mathcal{O}(\eta^2)$ for $\eta \ll 1$, then eliminating terms of order higher than linear in the equation governing $\bar{R}$ ((4) after simplifying for two cliques). We then expand this approximate equation for small $p$, retaining only lowest order terms. The resulting system is linear and can be solved exactly for the critical time $t_c$ at which $\bar{R}$ rises to $(1+x_0)/2$. The delay is simply the difference between that time and $t_{c_0} = -\ln(x_0/(1+x_0))/(2u-1)$, the critical time for the all-to-all $p = 1$ system.

Based on numerical work, the general behavior of this approximation—$d \propto -\ln p$—seems to remain valid even for large $p$, although the additive constant seems to change (see Figure S7). That is to be expected, since $d \to 0$ is required for $p \to 1$.

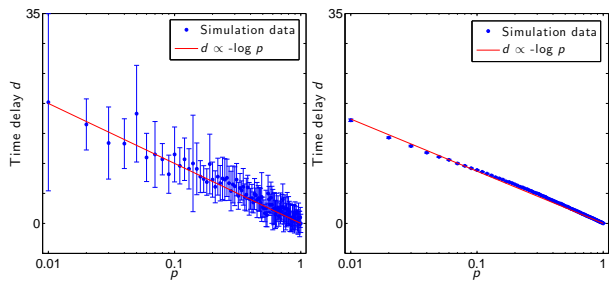

SUPPORTING FIG. S7. Variation of time delay $d$ with increasing inter-cluster connection probability $p$. Points and error bars indicate mean and standard deviation with 10 realizations. Lines show estimated logarithmic dependence. Left Panel: Discrete stochastic system (2). Right panel: Continuous deterministic system (4). For all simulations $N = 500$, $x(0) = 0.1$ and $u = 0.6$.

  


\end{document}